\documentstyle[12pt]{article}

\newcommand{\unit}[1]{\mbox{\ #1}}
\newcommand{\subs}[1]{\mbox{\scriptsize\it #1}}
\newcommand{\comment}[1]{}

\newcommand{\be}{\begin{equation}}
\newcommand{\ee}{\end{equation}}

\newcommand{\mcal}{{\cal M}}

\newcommand{\gp}{\gamma_{+}}
\newcommand{\faktor}{G_{F} \cos \theta_{c} e}
\newcommand{\Res}{\mbox{Res}}
\newcommand{\rp}{{\rho'}}
\newcommand{\rpp}{{\rho''}}

\newcommand{\br}{\mbox{BR}}
\newcommand{\strich}[1]{#1  \! \! \slash}

\begin{document}
\everymath={\displaystyle}
\thispagestyle{empty}
\vspace*{-2mm}
\thispagestyle{empty}
\noindent
\hfill TTP93--25\\
\mbox{}
 \hfill  July 1993  \\
\vspace{0.5cm}
\begin{center}
  \begin{Large}
  \begin{bf}
RADIATIVE CORRECTIONS\\
TO THE DECAY $\tau \to \pi (K) \nu_\tau$
   \\
  \end{bf}
  \end{Large}
  \vspace{0.8cm}
  \begin{large}
   Roger Decker and Markus Finkemeier\\[5mm]
    Institut f\"ur Theoretische Teilchenphysik\\
    Universit\"at Karlsruhe\\
    Kaiserstr. 12,    Postfach 6980\\[2mm]
    76128 Karlsruhe\\ Germany\\
  \end{large}
  \vspace{4.5cm}
  {\bf Abstract}
\end{center}
\begin{quotation}
\noindent
  We have calculated the $O(\alpha)$ radiative corrections to the
tau decay $\tau \to \pi (K) \nu$, taking into account both the point
meson contribution and the structure dependent radiation. We find
for the ratio
$\Gamma(\tau \to \pi \nu (\gamma)) / \Gamma(\pi \to \mu \nu (\gamma))$
a radiative correction of $+ 1.2 \%$ and for
$\Gamma(\tau \to K \nu(\gamma)) / \Gamma(K \to \mu \nu (\gamma))$
one of $+ 2.0 \%$.
We compare our results with an earlier estimation and with experimental data.
\end{quotation}
\newpage

%
%
%
Neglecting electromagnetic corrections, the decay rate $\Gamma(\tau \to
M \nu_\tau)$ (where $M = \pi$ or $K$) can easily be predicted from
$\Gamma(M \to \mu \nu_\mu)$. The relation at the order $O(\alpha^0)$
is:
\be \label{eqn1}
   R_{\tau/M}^{(0)} = \frac{\Gamma(\tau \to M \nu_\tau)}
   {\Gamma(M \to \mu \nu_\mu)} =
   \frac{1}{2}
   \frac{m_\tau^3}{m_M m_\mu^2}
   \frac{\left(1 - m_M^2 / m_\tau^2\right)^2}
   {\left(1 - m_\mu^2 / m_M^2 \right)^2}
\ee
Using current particle data
and the theoretical
prediction for $\tau \to e \nu \bar{\nu}$ (for details see below),
this yields the $O(\alpha^0)$ prediction
\be
  \left(
  \frac{\br(\tau \to \pi \nu) + \br(\tau \to K \nu)}
   {\br(\tau \to e \nu \bar{\nu})} \right)^{{(theo,0)}} =
   0.650
\ee
Recent experimental averages for these branching ratios have been
given in Ref.\ \cite{Cad93}, resulting in
\be \label{eqn2}
  \left(
  \frac{\br(\tau \to \pi \nu) + \br(\tau \to K \nu)}
   {\br(\tau \to e \nu \bar{\nu})} \right)^{{(expt)}} =
   0.675 \pm 0.015
\ee
So the $O(\alpha^0)$ theoretical prediction is off by $1.7$ standard
deviations. This discrepancy might either point to the importance of
the neglegted radiative corrections of $O(\alpha)$, or the experimental
numbers might simply not be quite correct. (Note that the new $\tau$ data
has greatly reduced the inconsistency between the values for the $\tau$ mass,
lifetime and its leptonic branching ratio, but there is still a hint that the
value given for the leptonic branching ratio might be  too small.)

Marciano and Sirlin \cite{Mar88} have given an estimate of the QED correction.
On the basis of the short distance behaviour of the weak interaction they
estimate the leading $O(\alpha \ln m_Z)$ corrections to both decays
as
\be \label{eqn3}
   R_{\tau/\pi}^{M.S.} =    R_{\tau/\pi}^{(0)}
    \frac{1 + 2 \alpha/\pi \ln(m_Z/m_\tau)}
   {1 + \frac{3}{2}(\alpha/\pi) \ln(m_Z/m_\pi)
      + \frac{1}{2}(\alpha/\pi) \ln (m_Z/m_\rho)}
\ee
(and similarly for the kaon mode, where $m_\pi \to m_K$),
which, using the current particle data, yields:
\be
  \left(
  \frac{\br(\tau \to \pi \nu) + \br(\tau \to K \nu)}
   {\br(\tau \to e \nu \bar{\nu})} \right)^{{(M.S.)}} =
   0.646
\ee
This deviates from the experimental value by two standard deviations. But
note that Eqn.\ (\ref{eqn3}) can be rewritten as
\be
   R_{\tau/\pi}^{M.S.} \approx    R_{\tau/\pi}^{(0)}
   \left[ 1 + \frac{\alpha}{2 \pi} \ln
    \left(
    \frac{m_\tau^4} {m_\pi^3 m_\rho }
    \right) \right]
\ee
The large mass $m_Z$ actually drops out in the ratio
and so the neglected terms
which are non-leading in $m_Z$ are of the same order of magnitude and must
be taken into account in order to get a reliable result.

We conclude that the increasing experimental accuracy asks for
a reconsideration of the $O(\alpha)$ corrections to these $\tau$ decays.
We have calculated these 
corrections by 
taking into account both the point meson contribution and the
structure dependent radiation.
In the usual manner \cite{Kin59} we replace
the vector-minus-axial vector current interaction
\be
   G_F \cos \theta_C f_\pi [\bar{\Psi}_l \gamma^\mu (1 - \gamma_5)
   \Psi_\nu] (i \partial_\mu - e A_\mu ) \Phi_\pi
\ee
(which yields the point meson contribution)
by the scalar-minus-pseudoscalar current interaction
\be
   G_F \cos \theta_C f_\pi m_l^{(0)}
   [\bar{\Psi}_l  (1 - \gamma_5) \Psi_\nu] \Phi_\pi
\ee
where $m_l^{(0)}$ denotes the bare lepton mass.

For the structure dependent radiation we use the ansatz of Ref.\
\cite{Dec93}, which we quote here for completeness (for details see
there):
\be
   \mcal_{SD}  =  \frac{\faktor}{\sqrt{2}} \left\{ i \epsilon_{\mu \nu \rho
      \sigma} L^\mu \epsilon^\nu k^\rho p^\sigma \frac{F_V(t)}{m_\pi}
      + \bar{u}(q) \gp \left[ (p \cdot k) \strich{\epsilon} -
      (\epsilon \cdot p) \strich{k} \right] u(s) \frac{F_A(t)}{m_\pi}
      \right\}
\ee
where the form factors are
\begin{eqnarray}
   F_A(t) & = & - 0.0116\, \Res_{a_1}(t)
\nonumber \\ \nonumber \\
   F_V(t) & = &
            \frac{- 0.0270}{1.085}
            \left[\Res_\rho(t) +
            0.136 \Res_\rp(t) - 0.051 \Res_\rpp(t) \right]
\end{eqnarray}
with normalized Breit-Wigner like resonance factors
\be
   \Res_X(t) = \frac{m_X^2}%
      {m_X^2 - t - i m_X \Gamma_{X}(t)}
\ee
with energy dependent widths, as described in the quoted reference.

We write the radiative corrected decay rates as
\be
   \Gamma(\mbox{channel}) =
   \Gamma^0(\mbox{channel}) \left[ 1 + \frac{\alpha}{2 \pi}
   \Bigg( B(\mbox{channel}) + B_{SD} (\mbox{channel}) \Bigg) \right]
\ee
where $B\mbox{(channel)}$ is the point meson contribution,
ie. the sum of virtual,
soft and hard brems\-strahlung, and $B_{SD}\mbox{(channel)}$ arises from
the structure dependent radiation (and its interference with the
point meson internal bremsstrahlung).

We find
\be
   B\Bigg(M \to l \nu_l (\gamma) \Bigg)  =
   - \frac{3}{2} \Delta
   + \frac{3}{2} \ln \frac{m_M^2}{\mu^2} + \frac{19}{4}
   - \frac{2}{3} \pi^2  + 6 \ln r_l
   + f(r_l)
\label{eqn16}
\ee 
and
\be \label{eqn17}
   B\Bigg(\tau \to M \nu_\tau (\gamma)\Bigg)
   =    - \frac{3}{2} \Delta
   + \frac{3}{2} \ln \frac{m_\tau^2}{\mu^2} + \frac{25}{4}
    - \frac{2}{3} \pi^2 + g(r_M)
\ee 
where
\begin{eqnarray}
   f(r_l) & = & 4 \left(\frac{1+r_l^2}{1-r_l^2} \ln r_l
   - 1 \right) \ln(1-r_l^2)
   - \frac{r_l^2 (8 - 5r_l^2)}{(1-r_l^2)^2}\ln r_l
\nonumber \\ \nonumber \\
   & & + 4 \frac{1 + r_l^2}{1-r_l^2} \mbox{Li}_2(r_l^2)
   - \frac{r_l^2}{1-r_l^2}
   \left( \frac{3}{2} + \frac{4}{3} \pi^2 \right)
\end{eqnarray}
and
\begin{eqnarray}
   g(r_M) & = & 4 \left( \frac{1+r_M^2}{1-r_M^2} \ln r_M - 1 \right)
                                       \ln(1-r_M^2)
   - \frac{r_M^2(2-5r_M^2)}{(1-r_M^2)^2} \ln r_M
\nonumber \\ \nonumber \\
   &  & + 4 \frac{1+r_M^2}{1-r_M^2} \mbox{Li}_2(r_M^2)
   + \left( \frac{3}{2} - \frac{4}{3} \pi^2 \right) \frac{r_M^2}{1-r_M^2}
\end{eqnarray}
In these formulae $\Delta$ denotes
\be
  \Delta = \frac{2}{\epsilon} - \gamma_{\subs{Euler}} +  \ln 4 \pi
\ee
where $4 - \epsilon$ is the number of space-time dimensions. $\mu$ is the
mass scale of dimensional regularization, $\mbox{Li}_2(x)$ is the dilogarithmic
function
\be
   \mbox{Li}_2(x) = - \int_0^x dt \, \frac{\ln(1-t)}{t}
\ee
and the mass ratios
$r_l$ and $r_M$ are defined by
\be
   r_l  =  \frac{m_l}{m_M} \mbox{\hspace*{2cm}\hspace*{2cm}}
   r_M  =  \frac{m_M}{m_\tau}
\ee
Note that $f(0) = g(0) = 0$, so from Eqn.\ (\ref{eqn16}) we find the well-known
mass singularity of the total radiative correction to pion decay as
$3 \alpha / \pi \ln (m_l)$, whereas in the radiative correction
to the tau decay the pion mass singularities cancel according to Eqn.
(\ref{eqn17}).
The zeroth order
pion decay rate is proportional to $m_\l^2$, whereas the tau decay rate
does not vanish for zero pion mass. So both the pion and tau results
agree with the Kinoshita-Lee-Nauenburg theorem on the cancelation of
mass singularities \cite{Kin59,Que92}.

For the structure dependent contribution numerical integration yields:
\begin{eqnarray}
    B_{SD} \Bigg(\pi \to \mu \nu_\mu (\gamma) \Bigg)  =  1.8 \cdot 10^{-5}
 & &
     B_{SD} \Bigg(\pi \to e \nu_e (\gamma) \Bigg)  =  0.66
\nonumber \\
    B_{SD} \Bigg(K \to \mu \nu_\mu (\gamma) \Bigg)  =  0.060
    \mbox{\hspace{0.6cm}}
\nonumber \\
    B_{SD} \Bigg(\tau \to \mu \nu_\mu (\gamma) \Bigg)  =  1.46
    \mbox{\hspace{0.9cm}}
  & &
    B_{SD} \Bigg(\tau \to K \nu_\tau (\gamma) \Bigg)  =  2.60
\label{eqn23}
\end{eqnarray}
We write the radiative corrections to the ratios as
\be
    R_{\dots} = R_{\dots}^0 ( 1 + \delta R_{\dots})
\ee
and find from the above
\begin{eqnarray}
     \delta R_{\tau/\pi} & = & + 1.05 \% + 0.17 \%= + 1.22 \%
\nonumber \\
     \delta R_{\tau/K} & = & + 1.67 \% + 0.29\% = + 1.96 \%
\end{eqnarray}
where the first numbers ($1.05$ and $1.67$, respectively) are the
point meson contributions and the second numbers give the structure
dependent part. So in both cases the model dependent 
contribution to the radiative correction is very small.
This is true for the integrated and inclusive
decay rate, whereas in certain regions of the differential decay distribution
the structure dependent part becomes dominant \cite{Dec93}.

Note that from Eqn.\ (\ref{eqn16}) and (\ref{eqn23}) we can also calculate
the radiative correction to the ratio of the electronic and muonic
decay modes of the pion
\be
   R_{e/\mu} = \frac{\Gamma(\pi \to e \nu_e (\gamma))}
   {\Gamma(\pi \to \mu \nu_\mu (\gamma))} =
   \frac{m_e^2}{m_\mu^2} \left(\frac{m_\pi^2 - m_e^2}{m_\pi^2 - m_\mu^2}
   \right)^2 \Bigg(1 + \delta R_{e/\mu}\Bigg)
\ee
as
\be
   \delta R_{e/\mu} = -3.93 \% + 0.08 \% = - 3.85 \%
\ee
where $-3.93 \%$ is the well known point pion correction \cite{Kin59}.

{}From the 1992 particle data book \cite{PDG92} we extract
\begin{eqnarray}
   \Gamma(\pi\to\mu \nu_\mu (\gamma)) & = & (2.528\pm 0.002)
   \cdot 10^{-14} \unit{MeV}
\nonumber \\
   \Gamma(K \to \mu \nu_\mu (\gamma)) & = & (3.41 \pm 0.02) \cdot 10^{-14}
   \unit{MeV}
\end{eqnarray}
and normalize to the standard model prediction for the
electronic decay mode of the $\tau$:
\begin{eqnarray}
   \Gamma(\tau \to e \nu \bar{\nu} (\gamma) & = &
   \frac{G_F^2 m_\tau^5}{192 \pi^3} \Bigg[ 1 - 8\frac{m_e^2}{m_\tau^2}
   + \frac{3}{5} \frac{m_\tau^2}{m_W^2} + \frac{\alpha}{2 \pi}
   \left( \frac{25}{4} - \pi^2 \right) \Bigg]
\nonumber \\ \nonumber \\
   & = & (4.033 \pm 0.005) \cdot 10^{-10} \unit{MeV}
\end{eqnarray}
using \cite{Cad93}
\be
   m_\tau = (1777.1 \pm 0.5) \unit{MeV}
\ee
And so we arrive at our predictions:
\begin{eqnarray}
  \left(
  \frac{\br(\tau \to \pi \nu(\gamma) )}
   {\br(\tau \to e \nu \bar{\nu}(\gamma) )} \right)^{{(theo,O(\alpha))}} & = &
   0.619 \pm 0.001
\nonumber \\
  \left(
  \frac{\br(\tau \to K \nu(\gamma) )}
   {\br(\tau \to e \nu \bar{\nu}(\gamma) )} \right)^{{(theo,O(\alpha))}} & = &
   0.0410 \pm 0.0002
\nonumber \\
  \left(
  \frac{\br(\tau \to \pi \nu(\gamma) ) + \br(\tau \to K \nu(\gamma) )}
   {\br(\tau \to e \nu \bar{\nu}(\gamma) )} \right)^{{(theo,O(\alpha))}} & = &
   0.660 \pm 0.001
\end{eqnarray}
The error bars given include only the uncertainties that result from
the uncertainties of the experimental input.
So our prediction agrees 
with the current experimental value
in Eqn.\ (\ref{eqn2}).

We finish by comparing our approach with that of Ref.\ \cite{Mar88}.
Their estimate is derived from the short distance
behaviour of the weak interaction, whereas we have used a point meson and the
hadronically dominated structure dependent radiation, ie. the long and
medium distance behaviour. Firstly this explains the fact that the logarithms
in $m_Z$, which our calculation gives when we put $
   \Delta = \ln (m_Z^2 / \mu^2)
$
do not agree with those in Eqn.\ (\ref{eqn3}). Furthermore we believe
that the short distance contributions are roughly the same for tau and
pion decay and so they drop out in ratios like $R_{\tau/\pi}$ at
least to a good approximation. And so our calculation should give
a rather reliable prediction, as it takes into account the dominant effects of
these decays.

\section*{Acknowlegdement}
One of us (M.F.) would like to thank A. Sirlin for an illuminating discussion.


\end{document}